\documentclass{aa} 
\usepackage[dvips]{graphics}
\begin{document}
%
%
\newcommand{\nc}{\newcommand}
\nc{\bea}{\begin{eqnarray}}
\nc{\eea}{\end{eqnarray}}
\nc{\beq}{\begin{equation}}
\nc{\eeq}{\end{equation}}
\nc{\ve}[1]{{\bf #1}}
\nc{\bP}{\mathbf{P}}
\nc{\bF}{\mathbf{F}}
\nc{\bZ}{\mathbf{Z}}
\nc{\bI}{\mathbf{I}}
\nc{\bA}{\mathbf{A}}
\nc{\bC}{\mathbf{C}}
\nc{\fk}{f_\mathrm{k}}
\nc{\etal}{{et al.}}


\title{A maximum likelihood approach to the destriping technique}

\author{E. Keih\"anen\inst{1,2} \and H. Kurki-Suonio\inst{1}
\and T. Poutanen\inst{2} \and D. Maino\inst{3} \and C. Burigana\inst{4}}

\offprints{H.~Kurki-Suonio, \email{hannu.kurki-suonio@helsinki.fi}}

\institute{University of Helsinki, Department of Physical Sciences,
P.O. Box 64, FIN-00014, Helsinki, Finland
\and Helsinki Institute of Physics, P.O. Box 64, FIN-00014,
Helsinki, Finland
\and Dipartimento di Fisica, Universit\'a di Milano,
Via Celoria 16, I-20131, Milano, Italy
\and IASF/CNR, Sezione di Bologna, Via Gobetti 101, I-40129, Bologna, Italy}

\date{ 
August 11, 2004}

\abstract{The destriping technique is a viable tool for removing different
kinds of systematic effects in CMB-related experiments. It has already been
proven to work for gain instabilities that produce the so-called $1/f$ noise
and periodic fluctuations due to e.g. thermal instability. Both effects, when
coupled to the observing strategy, result in stripes on the observed sky
region. Here we present a maximum-likelihood approach to this type of technique
and provide also a useful generalization. As a working case we consider a data
set similar to what the {\sc Planck} satellite will produce in its Low
Frequency Instrument (LFI). We compare our method to those presented in the
literature and find some improvement in performance. Our approach is also more
general and allows for different base functions to be used when fitting the
systematic effect under consideration. We study the effect of increasing the
number of these base functions on the quality of signal cleaning and
reconstruction. This study is related to {\sc Planck} LFI activities.

\keywords{methods: data analysis -- cosmology: cosmic microwave background} }

\authorrunning{Keih\"anen et al.}
\maketitle


\section{Introduction}

One of the major goals of cosmology is to determine the cosmological parameters
which describe the structure and evolution of the Universe. In this respect CMB
observations are a powerful tool directly probing the early phases of the
Universe. Recent results from the WMAP satellite (Bennett et al.
\cite{bennett03}) show that high accuracy in such measurements can be achieved
with an optimal choice of observing site (the second Lagrange point of the
Sun-Earth system, $L_2$, for good thermal and environmental stability), careful
instrument design and control of systematic effects. The last point is related
to the observing strategy adopted, which should be as redundant as possible,
with different measurements of the same sky region with different detectors and
on different time scales in order to properly control systematics.

Future space missions like {\sc
Planck}\footnote{http://astro.estec.esa.nl/SA-general/Projects/Planck/} which
are designed to have a signal-to-noise ratio of the order of few tens (far
larger than WMAP), require control of systematic effects at the $\mu$K level.
In this respect several techniques have been developed to treat systematics, to
detect and remove them in the best possible way. Burigana et al.
(\cite{Burigana99}), Delabrouille (\cite{Delabrouille98}) and Maino et al.
(\cite{Maino99}, \cite{Maino02}) have considered, in the context of the {\sc
Planck} mission, a simple destriping algorithm to remove systematics like the
$1/f^\alpha$ noise. Mennella et al. (\cite{Mennela02}) have instead considered
destriping when dealing with periodic fluctuations such as those induced by
thermal instabilities.

Destriping methods work  on time-ordered data (TOD) and produce TOD cleaned of
systematics. When TOD is cleaned it is possible to co-add observations on the
same region (pixel) of the sky to obtain a sky map which gives a visual
impression of the data. Although it is non-optimal, in the sense that it would
not necessarily produce the map with the minimum possible variance as instead
provided by the Generalized Least Square solution of the map-making problem
(see e.g. Natoli et al.~\cite{natoli01}), it provides a fast and accurate
map-making algorithm.  In addition, the analysis of TOD cleaned of systematics
is useful for several applications relevant for the {\sc Planck} data analysis
(e.g. in-flight main beam reconstruction (Burigana et al.~\cite{Bur02}) and
calibration (Bersanelli et al.~\cite{Ber97}, Piat et al. \cite{Piat03},
Cappellini et al.~\cite{Cap03}), time series analysis) and for the scientific
exploitation of {\sc Planck} data (e.g. source variability studies (Terenzi et
al.~\cite{Ter02})).

In this paper we consider the destriping technique in the light of
maximum-likelihood analysis and present a general formulation of the
destriping technique. We restrict our analysis to $1/f$ noise
fluctuations. They produce noise which is strongly correlated
in time and, when coupled with the observing strategy, will lead
to stripes in the final maps that would alter the
signal statistics. This is of extreme importance for the
CMB which is expected to be a Gaussian random field.

The basic idea in destriping is to model the noise in the TOD by a
linear combination of simple arithmetic functions, such as
polynomials or Fourier components. The amplitudes of these base
functions are determined taking advantage of the redundancy of the
scanning strategy for which the same points on the sky are
monitored several times during the mission. In its simplest form
destriping involves fitting uniform baselines, i.e. one baseline
for each elementary scanning period. In order to improve the
accuracy of the method, we present here the possibility of fitting
several components (base functions).

The destriping method of Burigana et al. (\cite{Burigana99}) and Maino
et al. (\cite{Maino99,Maino02}) differs from the destriping method
of Delabrouille (\cite{Delabrouille98}) in the weights they assign to
different map pixels based on the number of measurements falling on that
pixel. Our maximum-likelihood analysis presented in this paper leads
to a weighting scheme that differs from both of these. Therefore
we compare results obtained from all these three methods.

The paper is organized as follows. In Sect.~\ref{sec:ml} we present the
maximum-likelihood approach to the destriping technique, in Sect.~\ref{sec:uni}
we apply it in the case of uniform baselines, and in Sect.~\ref{sec:basef} we
generalize the discussion to arbitrary base functions. We present our
conclusions in Sect.~\ref{sec:conclu}.


\section{Destriping -- maximum likelihood approach}
\label{sec:ml}

\subsection{Maximum likelihood analysis}

In the following we present a maximum-likelihood based approach to
the destriping problem. We assume that data produced by a
generic detector at a given time $t$ could be written as:
\beq
    y_t = P_{tp} m_p + n_{t,{\rm corr}} + n_t \label{tod}
\eeq
where $m_p$ is the sky signal, assumed to be pixelized, $P_{tp}$ is the
pointing matrix, $p$ is the pixel index, $n_{t,{\rm corr}}$ is the
correlated noise component while $n_t$ is the white noise
component. The variance of the white noise component is
represented by a diagonal matrix $\mathbf{C}_n$ in the time
domain. Eq.(\ref{tod}) could be written in vector form as:
 \beq
    \ve{y} = \bP\ve{m} +\ve{n}_{\rm corr} +\ve{n}.
 \eeq

We model the correlated noise component of the TOD as follows. The TOD is
divided in to elementary scanning periods, which we shall here call ``rings''
(as appropriate for the {\sc Planck} scanning strategy). For each ring $j$ we
define a constant offset $a_j$, so that
 \beq
    y_t = P_{tp}m_p +F_{tj}a_j +n_t.
 \eeq
Here $F_{tj}$ equals unity if point $t$ lies on ring $j$.
We write this in matrix form as
 \beq
    \ve{y} = \bP\ve{m} +\bF\ve{a} +\ve{n}.
 \label{matrixform}
 \eeq

We treat both the map and the correlated noise component as
deterministic.
With these assumptions, we obtain the likelihood function
\beq
      \chi^2 = (\ve{y}-\bF\ve{a}-\bP\ve{m})^T
                \bC_n^{-1}(\ve{y}-\bF\ve{a}-\bP\ve{m}) .
      \label{chi1}
\eeq

If $N_t$ is the length (the number of samples) of the TOD stream,
$N_{\rm pix}$ is the number of pixels in the map, and $N_a$ is the
number of unknown amplitudes, then the sizes of the matrices are:
$[F]=(N_t,N_a)$, $[P]=(N_t,N_{\rm pix})$,
$[\mathbf{C}_n]=(N_t,N_t)$.

We now want to find the maximum likelihood solution for $\ve{a}$. We
need to minimize the function in Eq.~(\ref{chi1})
with respect to both of the unknown variables $\ve{m}$ and
$\ve{a}$.
First we find the minimum with respect to $\ve{m}$,
\beq
    \nabla_{m}\chi^2
    = -2\bP^T \bC_n^{-1}(\ve{y}-\bF\ve{a}-\bP\ve{m}) =0.
\eeq
From this we can solve the map $\ve{m}$,
\beq   \ve{m}= (\bP^T\bC_n^{-1}\bP)^{-1} \bP^T\bC_n^{-1}(\ve{y}-\bF\ve{a}) \, .
       \label{mlmap}
\eeq

Substituting Eq.~(\ref{mlmap}) back into Eq.~(\ref{chi1}) we obtain
\beq
    \chi^2 = (\ve{y}-\bF\ve{a})^T \bZ^T\bC_n^{-1}\bZ(\ve{y}-\bF\ve{a}), \label{chin}
\eeq
where
\beq
      \bZ = \bI-\bP(\bP^T\bC_n^{-1}\bP)^{-1} \bP^T\bC_n^{-1} \, .
\eeq
Here $\bI$ denotes the unit matrix.

The next step would be to minimize $\chi^2$ with respect to $\ve{a}$,
\beq
    \nabla_a\chi^2 = -2\bF^T \bC_n^{-1}\bZ(\ve{y}-\bF\ve{a}) = 0.
\eeq
The minimum is given by
\beq
    \bF^T\bC_n^{-1}\bZ\bF\ve{a} = \bF^T\bC_n^{-1}\bZ\ve{y}.  \label{linc}
\eeq
Here we have used the property $\bZ^T\bC_n^{-1}\bZ=\bC_n^{-1}\bZ$.

We assume from now on $\bC_n=diag(\sigma^2)$.
With this simplification, the minimum of (\ref{chin}) is given by
\beq
    \bF^T\bZ\bF\ve{a} = \bF^T\bZ\ve{y}.  \label{lin1}
\eeq
where
 \beq
      \bZ = \bI-\bP(\bP^T\bP)^{-1} \bP^T \, .
 \label{zdef}
 \eeq
The effect of $\bZ$ acting on a TOD is to subtract from each sample the average
of all samples hitting the same pixel. The solution to Eq.~(\ref{lin1}) is not
unique. We may add an arbitrary constant to $\ve{a}$ without changing the value
of $\chi^2$. This is equivalent to varying the monopole component of the CMB
map, and is irrelevant for anisotropy measurements. To remove this ambiguity we
require that the sum of baselines is zero, $\ve{a}^T\ve{1}=0$. Here $\ve{1}$ is
a vector with all elements equal to 1. This is equivalent to adding term
$\ve{1}\ve{1}^T\ve{a}$ to the left-hand side of Eq.(\ref{lin1}).

The solution is now given by
 \beq
    \ve{a} = \lbrack \bF^T\bZ\bF +\ve{1}\ve{1}^T\rbrack^{-1} \bF^T \bZ\ve{y}.
      \label{asolu2}
 \eeq
Matrix $\bF^T\bZ\bF +\ve{1}\ve{1}^T$, unlike $\bF^T\bZ\bF$, is non-singular,
provided that there are enough intersection points between the rings.
$\ve{1}\ve{1}^T$ denotes a matrix with all elements equal to one. When the
amplitude vector $\ve{a}$ has been determined, the CMB map can be computed
according to Eq.~(\ref{mlmap}). These equations are the main theoretical result
of this paper.

\subsection{Pointing and beam shape}

The matrix $P$ spreads the map $\ve{m}$ into TOD. In principle, the beam shape
and profile can be incorporated in $\bP$. The elements of $\bP$ then determine
the weights that different pixels contribute to a given measurement. In this
way beams with arbitrary shapes and profiles can be treated. In this case the
matrix $\bP^T\bP$ is non-diagonal. Due to its large size, in {\sc Planck} type
of missions its inversion would present a major computational burden.

A simpler approach is to consider each sample in the TOD to
represent the temperature of the pixel at the center of the beam.
Then $\bP$ takes a particularly simple form, consisting of ones and
zeros for a full-power measurement like {\sc Planck}. Each row
contains one non-zero element identifying the pixel on which the
corresponding measurement falls. Matrix $\bP^T\bP$ becomes diagonal,
the diagonal elements giving the number of hits on each pixel.  We
follow this approach from here on.

\subsection{Comparison with earlier work}
\label{comparison}

Equation(\ref{chin}) can be put into form
\beq
   \chi_n^2 = \sum_p\frac{\sum_{ik,jl}
              (a_i-a_j-y_{ik}+y_{jl})^2 d^p_{ik}d^p_{jl}}
                 {2\sigma^2\sum_{ik}d^p_{ik}}   \label{chin_index}
\eeq
where index $p$ labels pixels, $i,j$ scanning rings, and
$k,l$ samples on a given ring. A combined index $ik$ or $jl$
identifies a measurement. We define the symbol $d^p_{ik}$ so that
$d^p_{ik}=1$ if measurement $ik$ falls into pixel $p$, otherwise
$d^p_{ik}=0$. Due to the factors $d^p_{ik}$ and $d^p_{jl}$ in
the numerator of Eq.~(\ref{chin_index}) only those pixels
$p$ contribute to the pixel sum which lie on two or more
scanning rings, and the sum $\sum_{ik,jl}$ is equal to
2 times the sum over all {\em pairs} of measurements falling on pixel $p$.

Eq.~(\ref{chin_index}) can be compared to Eq.~(2) of Maino et al.
(\cite{Maino99}) or to Eq.~(10) of Burigana et al. (\cite{Burigana99}). The
formulae differ in that in Eq.~(\ref{chin_index}) $\chi_n^2$ has in the
denominator the term $n_p=\sum_{ik}d^p_{ik}$, which gives the total number of
hits in pixel $p$.

Delabrouille (\cite{Delabrouille98}) gives the general form
\beq
   S = \sum_{p\in {\rm sky}}\sum_{\rm pairs}
         w(p,ik,jl)(y_{ik}-y_{jl}-a_i+a_j)^2 \label{dela}
\eeq for the function $S$ to be minimized. Here the second sum refers to all
pairs $(ik,jl)$ that can be formed of the measurements falling onto pixel $p$,
and $w$ is a weight function to be chosen. Based on the fact that pixel $p$
contributes $n_p(n_p-1)/2$ pairs, Delabrouille suggests choosing
$w\propto1/(n_p-1)$. The result of Maino et al. (\cite{Maino99}) corresponds to
$w=\mbox{const.}$ and our new result, Eq.~(\ref{chin_index}), to $w=1/n_p$.

\subsection{Circular scanning}

In the nominal scanning strategy of {\sc Planck} the spin axis follows the
ecliptic plane. The spin axis is kept anti-solar
by repointing it by $2.5'$ every hour. The spacecraft rotates around the
spin axis at a rate of 1 rpm.
During one hour {\sc Planck} scans the same circle on the sky 60 times.
As the sky signal is almost time-independent, the data can be coadded to reduce
the length of the data stream by a factor of 60.  We call this set of 60
circles, and the corresponding segment of the coadded TOD a ``ring''.
The crossing points of the rings are important calibration
points, which allow for the removal of the correlated noise component from
the TOD.

The opening angle of the scanning circle varies between 80-90 degrees,
depending on the location of the detector on the focal plane.
The sampling frequency for the 100 GHz LFI receiver is 108.3 Hz.
The instrument then collects 6498 temperature values, or ``samples'',
at each rotation of the spacecraft, corresponding roughly to
a $3'$ shift between successive samples. A total of 8766
rings builds up one year of observations.

In reality, the angular velocity of the rotation does not stay
exactly constant, especially immediately after repointing, and the
samples from different circles of the same ring are shifted in
position. This will probably require discarding the first few
circles of each ring, and resampling or phase binning the rest
before performing the coaddition. During the ring the spin axis of
the satellite will follow a nutation ellipse with maximum
amplitude of 1.5$'$ at the end of the nutation damping phase (van
Leeuwen et al. \cite{vanL02}). This may degrade slightly the
performance of destriping. We ignore these complications in this
paper.

The destriping technique applies particularly well to a {\sc Planck}-like
measurement pattern resulting from the coadding of scanning circles into
rings which breaks the stationarity of the data.
The $1/f^\alpha$ noise component in the coadded TOD
is well presented by a piecewise defined function, where each piece
consists of a linear combination of a few base functions.

\section{Uniform baselines}
\label{sec:uni}

\subsection{Simulation results}
\label{sec:unisim}

We have carried out simulations of the Planck LFI 100 GHz detector. The
underlying CMB map was created by the Synfast code of the HEALPix
package\footnote{http://www.eso.org/science/healpix}, starting from the CMB
anisotropy angular power spectrum computed with the CMBFAST
code\footnote{http://physics.nyu.edu/matiasz/CMBFAST/cmbfast.html}
 (see Seljak \& Zaldarriaga
 \cite{cmbfast96}, and references therein)
using the cosmological parameters $\Omega_{\rm tot}=1.00$,
$\Omega_\Lambda=0.7$, $\Omega_\mathrm{b}h^2=0.02$, $h=0.7$, $n=1.00$, and
$\tau=0.0$. We created the input map with HEALPix resolution
$N_\mathrm{side}=1024$ and with a symmetric Gaussian beam with a full width at
half maximum (FWHM) of $10'$. We then formed the signal TOD by picking
temperatures from this map. All our output maps have the resolution parameter
$N_\mathrm{side}$=512, corresponding to an angular resolution $7'$.

The scanning pattern corresponds to the nominal {\sc Planck}
scanning strategy of the 100 GHz LFI detector number
10.\footnote{Simulation Software is part of the Level S of the
{\sc Planck} DPCs and is available for {\sc Planck} collaboration
at http://planck.mpa-garching.mpg.de.} The angle between the
satellite spin axis and the optical axis of the telescope is
$85\degr$. The beam center is pointing towards $(\theta,\phi) =
(3\fdg737,126\fdg228)$. Here $\theta$ is the angle from the
optical axis and $\phi$ is an angle counted clockwise from the
axis pointing from the center of the focal plane towards the
satellite spin axis. We assumed no spin axis precession. Our
simulated data set consists of 5040 scanning rings, corresponding
to 7 months of measurement time. The TOD stream contains 6498
samples on each ring, corresponding to a sampling frequency of
$f_\mathrm{s}=108.3$ Hz. The sky coverage is 98.5\%.

In our simulations we have assumed a symmetric Gaussian beam, and convolved the
input map with the beam.

The rings cross at points which are mostly concentrated near
the ecliptic poles. We count as crossing points all points where two
measurements on different rings fall on the same pixel.
Since our pixel resolution ($7'$) exceeds the repointing angle of the
spin axis ($2.5'$), the crossing points include cases where two nearby
rings pass parallel through the same pixel, without
actually crossing each other.

We used the Stochastic Differential Equation (SDE) technique to create the
instrument noise stream, which we added to the signal TOD.\footnote{SDE is one
of the two methods in the Planck Level S pipeline for producing simulated
instrument noise. The method was implemented by B.~Wandelt and K.~G\'orski and
modified by E.~Keih\"{a}nen.} We generated noise with the power spectrum
 \beq
   P(f) = \left(1+\frac{\fk}{f}\right)\frac{\sigma^2}{f_\mathrm{s}},
          \quad(f>f_\mathrm{min})
 \eeq
with parameters $\sigma=4800\mu\mbox{K}$ (CMB temperature scale),
$\fk=0.1\mathrm{Hz}$, and $f_\mathrm{min}=10^{-6}\mbox{Hz}$. The parameter
$\fk$ is called the knee frequency. The noise level $4800\mu\mbox{K}$
corresponds to the estimated white noise level of one $100\mbox{GHz}$ LFI
detector.

\begin{figure}[tbh]
\center{\resizebox{\hsize}{!}{\includegraphics{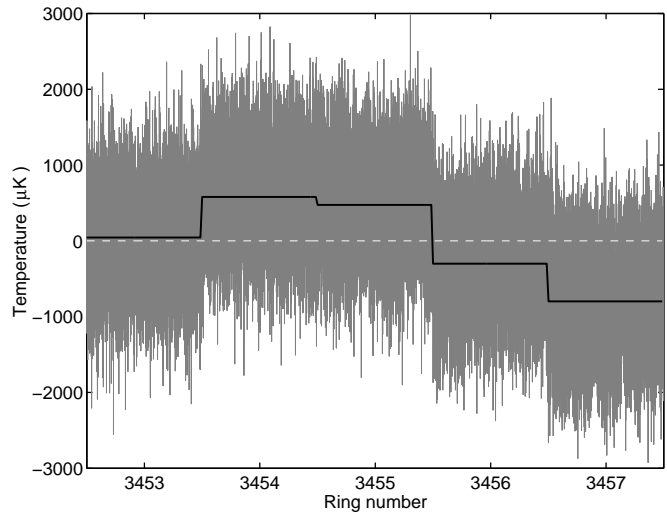}}}
 \caption{A 5 hour segment of the noise TOD after coadding (grey) and the
baselines (black) fitted to it. For this figure the 5-hour average is
subtracted also, to center the figure at 0 $\mu$K. The baselines were obtained
using the weight function $w=1/n_p$. The difference between baselines obtained
with different methods would be too small to show up clearly in this figure.}
 \label{fig:noise_illustration}
\end{figure}

\begin{figure}[tbh]
 \center{\resizebox{\hsize}{!}{\includegraphics{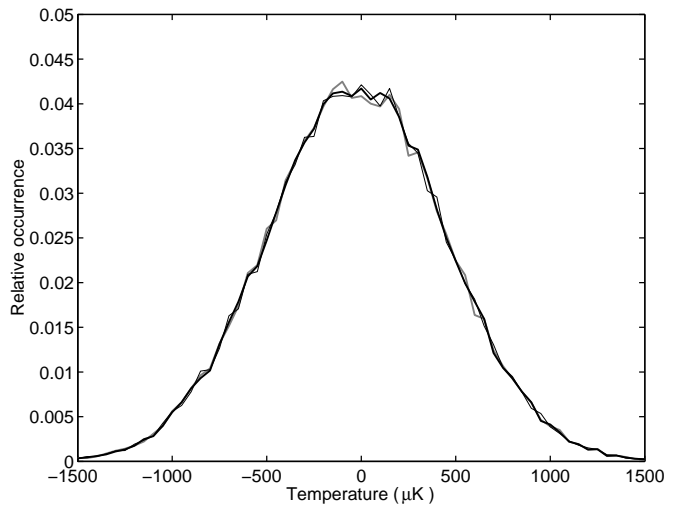}}}
 \caption{The histogram (in bins of 50 $\mu$K) of the baselines from a set
of 10 simulated 7-month noise TODs. The three curves correspond to two
different weight functions (see Sect.~\ref{sec:compweight}) used to determine
the baselines: $w=1/n_p$ (thick black) and $w = 1$ (thin black), and to the
reference baselines, i.e., noise averages over the one hour ring (grey).}
 \label{fig:uni_w_histogram}
 \end{figure}

Fig.~\ref{fig:noise_illustration} shows a five-hour section of a coadded noise
TOD and the baselines fitted to it.  Fig.~\ref{fig:uni_w_histogram} shows the
baseline distribution from a set of 10 simulated 7-month noise TODs.

There was no foreground included in the simulations presented in this paper,
but we have also verified our destriping codes on simulated data sets with
foreground. We found that the foreground has an insignificant impact on the
baselines determined by the destriping, in agreement with the discussion in
Maino et al. (\cite{Maino02}). The quality of destriping is also almost
independent of the impact of another class of instrumental systematic effects,
main beam distortions and straylight, as the temperature differences at
crossing pixels are dominated by the noise and only minimally affected by the
spurious signals ($\approx \mu$K) introduced by optical distortions (see e.g.
Burigana et al. 2001).

\begin{figure}[tbh]
\center{\resizebox{\hsize}{!}{\includegraphics{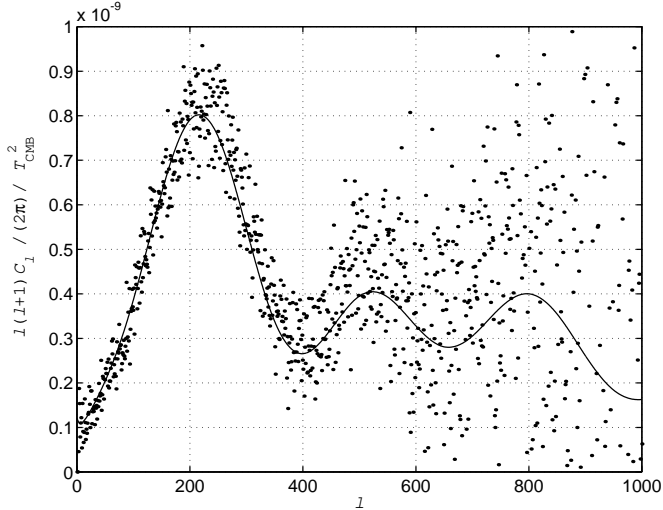}}}
\caption{The input $C_\ell$ spectrum and the spectrum computed
from the CMB map. The latter has been corrected for beam and pixel
convolution, and for white noise level. Destriping was done
fitting uniform baselines only.} \label{fig:cls1}
\end{figure}

Fig.~\ref{fig:cls1} shows the input $C_\ell$ spectrum and the spectrum derived
from the simulated TOD after destriping. We used the Anafast code of the
HEALPix package to compute the $C_\ell$ spectrum of the destriped map. We
subtracted from the derived spectrum an estimate of the noise level $C_{\rm
noise}=0.197\mu{\rm K}^2$ (estimated as the average of $C_\ell$ over
$\ell=980\ldots1000$) and corrected the spectrum for the beam shape convolution
and pixel convolution. The angular spectrum shown is thus $\tilde
C_\ell=(C_\ell-C_{\rm noise})/(B^2_\ell h^2_\ell)$, where
$B_\ell=\exp(-\sigma_b^2\ell(\ell+1)/2)$, with $\sigma_b=10'/\sqrt{8\ln(2)}$,
is the beam convolution function corresponding to the assumed beam width of
$10'$ (FWHM), and $h_\ell$ is the pixel convolution function (provided by the
HEALPix package).

\begin{figure}[tbh]
\center{\resizebox{\hsize}{!}{\includegraphics{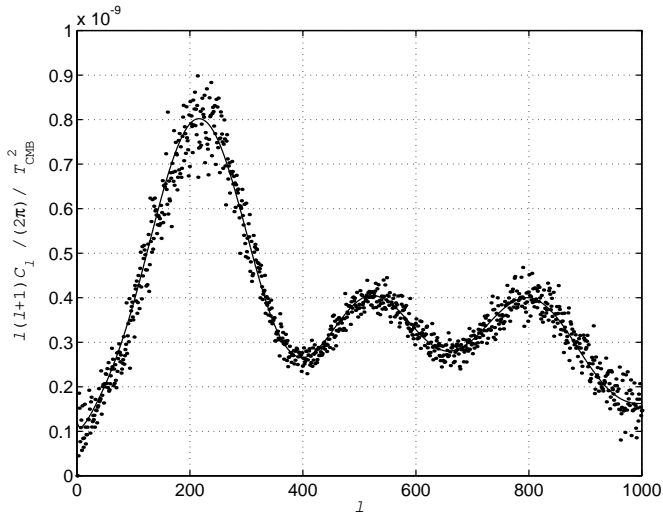}}}
\caption{Same as Fig. \ref{fig:cls1}, but for noise level lowered
by factor $1/\sqrt{24}$. This simulates the effect of combining 24
detectors.} \label{fig:cls2}
\end{figure}

Fig.~\ref{fig:cls2} shows the same for a noise level reduced by the factor
$\sqrt{24}$, corresponding to the combination of 24 detectors. Here the
subtracted noise level was $C_{\rm noise}=0.197\mu{\rm K}^2/24=0.0082\mu{\rm
K}^2$.

Note that Figs.~\ref{fig:cls1} and \ref{fig:cls2} are for illustration only,
as this paper does not address the full CMB $C_\ell$ estimation problem, and thus
we have just used the above crude estimate for $C_\mathrm{noise}$.

\subsection{Comparison of different weighting schemes: maps and angular power spectra}
\label{sec:compweight}

We have written a destriping code which allows us to compare the different
weight functions discussed in Sect.~\ref{comparison}. We use the Cholesky
decomposition technique to solve the set of linear equations.

We have chosen the root-mean-square (rms) value of the residual noise map
(see below) pixels as a
figure of merit that we use to compare different destriping methods.
This map rms is related to the $C_\ell$ spectrum through the relation
 \beq
    {\rm rms}^2 = \sum_\ell\frac{2\ell+1}{4\pi f_{\rm sky}}C_\ell. \label{clrms}
 \eeq
The map rms squared is thus a weighted sum of the angular spectrum,
 with high weight on high multipoles.
The sky coverage $f_{\rm sky}=0.985$ enters here because we have
computed our rms values over the visited pixels only.
When computing $C_\ell$ spectra, we have set $T=0$ in the remaining pixels.

We compute the residual noise map by taking
the difference between the destriped map and the noise-free reference
map and subtracting the monopole component.
Note that because of the incomplete sky coverage, removing the monopole
affects the $C_\ell$ spectrum at all $\ell$ (not only $\ell=0$).
The reference map is computed by coadding the pure signal TOD into
a map of resolution $N_{\rm side}$=512.
The expected contribution from white noise to the residual map rms
is 220.95 $\mu$K.

While the rms of the residual noise map is a natural measure of
the CMB map quality,
the main scientific interest is perhaps not in the CMB map
itself, but rather in its angular power spectrum $C_\ell$.  It is therefore
of interest to see the impact of the destriping methods on the different
parts of the $C_\ell$ spectrum.  The map rms is dominated by the high $\ell$
part, and does not reveal the difference in performance
of the various methods in the low $\ell$ part.
Thus we have computed the angular power
spectra $C_\ell$ of the residual noise maps.

Because of the random nature of the noise, the result of a comparison between
methods may vary from one noise realization to another. We therefore performed
destriping 10 times, with different realizations of instrument noise. The
underlying CMB map was kept the same. We then used the average of the 10
residual map rms values and the residual noise map $C_\ell$ spectra to compare
the methods. We also calculated the standard deviation (std) of the 10 map rms
and $C_\ell$ values, to see whether the differences between the methods were
statistically significant.
Thus this average rms approximates the expectation value for the rms with an
accuracy of $\mbox{std}/\sqrt{10}$.  However, the std itself tells us how much
we can expect the residual map rms for a single realization to deviate from
this average value.

\begin{figure}[tbh]
\center{\resizebox{\hsize}{!}{\includegraphics{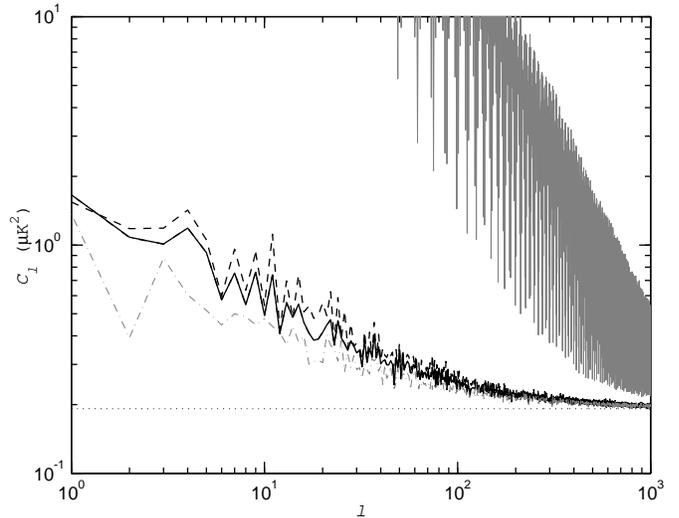}}} \caption{The
$C_\ell$ spectrum of the residual noise map, for different choices of the
weight function $w$. The spectra are averages over 10 realizations of noise.
Only uniform baselines are fitted. The lower {\it solid} line corresponds to
the choice $w=1/n_p$ for the weight function in Eq.~(\ref{dela}) and the {\it
dashed} line to $w=\mbox{const}$. The difference between weight functions
$w=1/n_p$ and $w=1/(n_p-1)$ is too small to show on this plot. The difference
between them is plotted in Fig.~\ref{fig:diffweight} (lower panel). The upper
{\it solid} (gray) line shows the spectrum of a naive coadded map (no
destriping). The {\it dash-dotted} (gray) line shows the ideal reference
spectrum, computed by removing the reference baselines. The corresponding map
rms values are shown in Table 1.  The {\it dotted} line shows the theoretical
white noise level $0.192\mu{\rm K}^2$.} \label{fig:clnweight}
\end{figure}

Table~1 presents a comparison between different weight functions discussed in
Sect.~\ref{comparison}. The corresponding $C_\ell$ spectra are shown in
Fig.~\ref{fig:clnweight}. Since from our simulations we have the noise streams
available separately, we also computed reference baselines as the average of
the noise stream over each ring. This reference baseline can be thought of as
the ``true'' baseline of the noise.  For comparison, we then subtracted the
reference baselines from the TOD and computed the rms of the resulting map.
This represents an ideal situation, where we could determine the baselines
exactly. The reference rms is given on the last line of Table 1. Actual
residual noise rms values are always larger.

Weight functions $w=1/n_p$ and $w=1/(n_p-1)$ in Eq.~(\ref{dela}) give similar
results, due to the fact that for most pixels $n_p\gg1$. Weight $w=1/n_p$
suggested by our maximum likelihood analysis is clearly superior to $w=1$.
However, the weight $w=1/(n_p-1)$ gives an even smaller rms, although the
difference is very small. The difference of the rms between $w=1$ and $w=1/n_p$
is significant because the difference is about 10 times larger than the
respective std of the rms.  Although the difference between $w=1/n_p$ and
$w=1/(n_p-1)$ is much less than the std between different noise realizations,
it was in the same direction in each realization. Note that we were able to
measure this small difference only because we used the same set of random seeds
for all weighting schemes.


\begin{table}
 \caption[a]{\protect\small Average (avg) rms and std of rms of the residual
noise map for different weight functions. The average and std are taken over 10
noise realizations. The corresponding $C_\ell$ spectra are shown in Fig.
\ref{fig:clnweight}. The last line gives the reference rms, which would be
reached if one could determine the baselines exactly.  The differences between
the rms are significant only if they are larger than the std.  We show extra
digits for the rms in Tables 1 and 2 to show the systematic (but insignificant)
difference (see text) between the  $w=1/(n_p-1)$ and $w=1/n_p$ cases.}
 \begin{center}
\begin{tabular}{lllll}
\hline\hline
Weight         & avg rms$/\mu$K & std of rms$/\mu$K  \\
\hline
$w=1  $        &  225.1619 & 0.072 \\
$w=1/(n_p-1)$  &  224.4332 & 0.073 \\
$w=1/n_p$      &  224.4443 & 0.073 \\
ref.           &  224.1170 & 0.075 \\
\hline
\end{tabular}
\end{center}
\label{tab:rms1}
\end{table}


\begin{table}
\caption[a]{\protect\small Average rms and std of rms of the
residual noise map, for a simplified noise model.
The noise consists of uniform baselines + white noise. }
\begin{center}
\begin{tabular}{lll}
\hline\hline
Weight         & avg rms$/\mu$K & std of rms$/\mu$K \\
\hline
$w=1  $        &  221.4816 & 0.069 \\
$w=1/(n_p-1)$  &  221.1041 & 0.072 \\
$w=1/n_p$      &  221.1039 & 0.072 \\
ref.           &  220.9593 & 0.073 \\
\hline
\end{tabular}
\end{center}
\label{tab:std2}
\end{table}

\begin{figure}[tbh]
\center{\resizebox{\hsize}{!}{\includegraphics{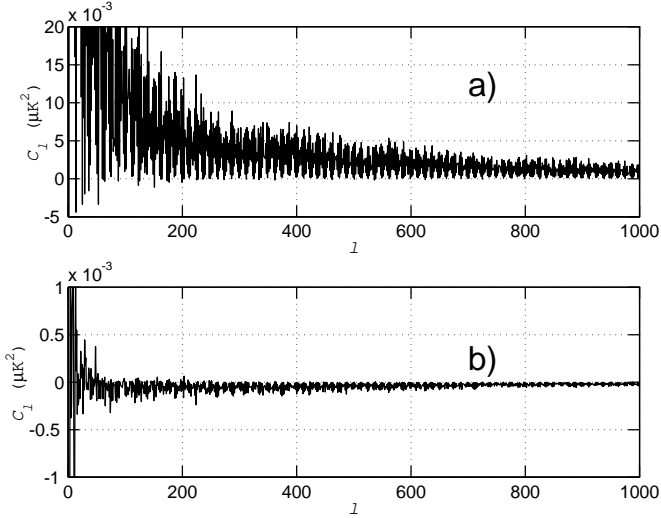}}}
\caption{Differences between the $C_\ell$ spectra shown in
Fig.~\ref{fig:clnweight}. {\bf a)} The difference
$C_\ell(w=\mbox{const}) - C_\ell(w=1/n_p)$. {\bf b)} The
difference $C_\ell(w=1/(n_p-1)) - C_\ell(w=1/n_p)$. Note the much
expanded vertical scale in Fig.~\ref{fig:diffweight}b. }
\label{fig:diffweight}
\end{figure}

Since the difference between the weight functions $w=1/n_p$ and $w=1/(n_p-1)$
is so small that it does not show up in Fig.~\ref{fig:clnweight}, we show
just the differences in Fig.~\ref{fig:diffweight}.

\begin{figure}[tbh]
\center{\resizebox{\hsize}{!}{\includegraphics{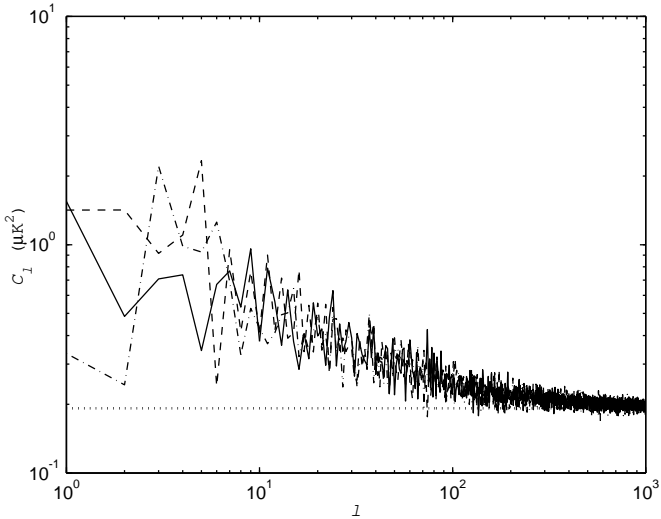}}}
\caption{Three examples of individual $C_\ell$ spectra, from
different noise realizations, after destriping. Only uniform
baselines are fitted.  The average of 10 such spectra is shown in
Fig.~\ref{fig:clnweight} (the solid line there).}
\label{fig:samples}
\end{figure}

It is well known that maximum-likelihood analysis should provide the
minimum-variance solution. Therefore it may at first seem surprising that the
maximum-likelihood based weight function did not give the best results.
However, the maximum-likelihood solution is the optimal one only if the model
used corresponds to reality. Here we have modelled the $1/f$ noise component in
the TOD by a uniform baseline, which is a simplified model. Further, we have
assumed that the baselines are independent from ring to ring. The reason for
the maximum-likelihood solution not giving the best result is that the noise
model used in the analysis does not exactly correspond to the actual noise
properties.

To verify this, we re-generated our input noise in a way that
better corresponds to the model assumed in the analysis.
We generated the $1/f$ noise in the usual way,
but then, for each ring, we took the average over the ring,
and replaced the original $1/f$ contribution to the ring with this
average value, on top of which we added white noise.
This way we obtained noise which still has a realistic
correlation between scanning rings, but consists of baselines + white
noise only.
The results are shown in Table 2. Now the maximum-likelihood based
weight function gives the smallest variance map.

Thus it seems that the slightly better performance of the Delabrouille
weighting scheme ($w = 1/(n_p-1)$) is related to the effect of that part
of the correlated noise which deviates from uniform baselines.

The scatter in the individual residual noise map $C_l$ values from one noise
realization to another was larger than the difference between the methods.
(See Fig.~\ref{fig:samples} for the  $C_\ell$ spectra from the first three
noise realizations, using our weighting scheme, $w=1/n_p$.)  The difference
between $w = 1$ and $w = 1/n_p$ becomes however statistically significant when
the $C_\ell$ are binned into larger bins.

%


\section{Increasing the number of base functions}
\label{sec:basef}

As shown by Delabrouille (\cite{Delabrouille98}) and Maino et al.
(\cite{Maino99,Maino02}), a simple model with
uniform baselines and white noise models quite well the coadded noise
stream.
One can try to further improve the performance
of destriping by introducing more base functions.
Delabrouille (\cite{Delabrouille98}) found that the addition of a small
number of base functions improved the performance of destriping,
whereas Maino et al. (\cite{Maino02}) found no significant improvement.
The difference between these results
could be due to the different noise spectra considered:
$1/f$ noise for Maino et al. (\cite{Maino02}) and $1/f^2$ for
Delabrouille (\cite{Delabrouille98}).

We generalize the discussion of Sect. \ref{sec:ml} to include arbitrary base
functions. We model the correlated noise component of the TOD by a linear
combination of base functions as $\ve{n}_{\rm corr}=\bF\ve{a}$. Here $\bF$ is a
matrix, whose columns contain the base functions, and $\ve{a}$ is a vector
containing their (unknown) amplitudes. It is convenient to select an orthogonal
set of the base functions, so that $\bF^T\bF$ is diagonal. Eqs.
(\ref{chi1})--(\ref{zdef}) hold as such for general baselines.

We have studied two sets of base functions:
Fourier components and Legendre polynomials.
Both form an orthogonal set.

\subsection{Simulation results}

The solution of the general destriping problem involves the solution
of a large linear system of equations
\beq
   \bF^T\bZ\bF\ve{a} = \bF^T\bZ\ve{y}. \label{lin3}
\eeq Matrix $\bA\equiv\bF^T\bZ\bF$ becomes very large if several base functions
are fitted. We use the iterative conjugate gradient method (see, e.g., Press et
al. \cite{NumRes}) to solve the system. The conjugate gradient method only
requires multiplication by matrix $\bA$. That can easily be done
algorithmically, without actually storing the full matrix $\bA$ at any one
time. The conjugate gradient method allows us to perform the destriping in a
relatively small memory space.

Note that we do not need to add the term $\ve{1}\ve{1}^T$, since the conjugate
gradient method has the property that, when solving system $\bA\ve{x}=\ve{b}$,
it automatically finds the solution for which $\ve{x}_0^T\ve{x}=0$, if
iteration is started with $\ve{x}=0$, and $\bA\ve{x}_0=0$ and
$\ve{b}^T\ve{x}_0=0$.  (The amplitude vector which gives unit amplitudes to the
uniform baselines and zero amplitudes to other baselines is an example of such
a vector $\ve{x}_0$, so the average of the uniform baselines is set to zero.
The case of other possible vectors in the null space of $\bA$ is discussed
further below.)

We compare four sets of functions:
\newline 1. (``Un.'') Uniform baselines only.
\newline 2. (``F1'')
Uniform baseline + first Fourier frequency, which gives three
components for each ring: constant, $\sin(2\pi f_{\rm sc}t)$, and $\cos(2\pi
f_{\rm sc}t)$ ($q=0,1,2$). Here $f_{\rm sc}=1/$(60 s) is the scanning
frequency.
\newline 3. (``L1'') Uniform baseline + 1st (linear) Legendre
polynomial.
\newline 4. (``L2'') Uniform baseline + 1st and 2nd
Legendre polynomials.

The first Legendre polynomials are
$L(x,0)=1$, $L(x,1)=x$, and $L(x,2)=\frac12(3x^2-1)$,
for $x\in[-1,1]$.

Again we averaged the residual noise $C_\ell$ spectra and the
residual noise map rms over 10 noise realizations.

\begin{figure}[tbh]
\center{\resizebox{\hsize}{!}{\includegraphics{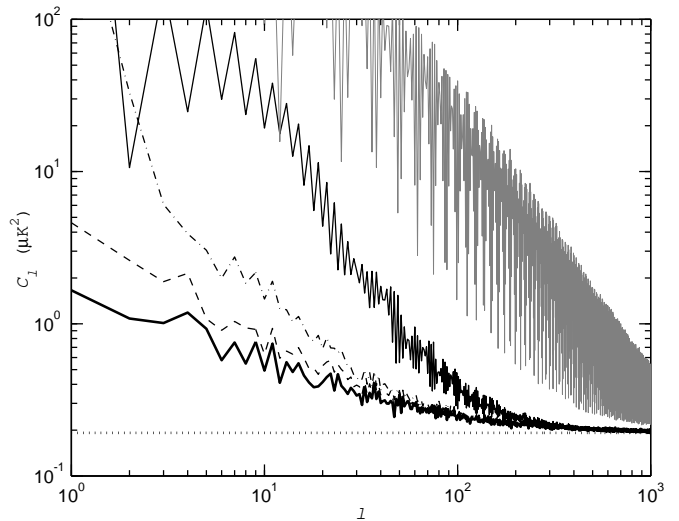}}}
\caption{Average $C_\ell$ spectrum of the residual noise map, for
different choices of the base functions. {\it Thick solid} line:
uniform baselines. {\it Thin solid} lower line: three Fourier
components ($q=0,1,2$). {\it Dashed} and {\it dot-dashed} lines:
Legendre polynomials up to 1st and 2nd order.
The upper {\it solid} (gray) line shows the
spectrum of a naive coadded map (no destriping). The spectra were
averaged over 10 realizations of noise.} \label{fig:clnd0}
\end{figure}

Table 3 and Fig. \ref{fig:clnd0} present
our results of fitting several base functions.
We find no improvement in the map rms.

The last column of Table 3 gives a reference rms, which was computed as
follows. We defined the reference amplitude vector as
$\ve{a}_0=(\bF^T\bF)^{-1}\bF^T\ve{n}$, where $\ve{n}$ is the pure noise stream
(i.e. $\ve{n}_{\rm corr} +\ve{n}$ of Eq.~(\ref{matrixform}). We coadded a map
from the TOD stream, from which we had removed the reference base functions,
$\ve{y}-F\ve{a}_0$, and computed the rms of the residual noise map. Fourier
components give the lowest reference rms, showing that Fourier components model
the noise best of our base function sets. However, when we fit the data,
Fourier components give the poorest results, as the first column of Table 3
shows. The worst of the 10 runs gave a rms of 452 $\mu$K. The code took a long
time to converge, and the final maps contained a very strong and obviously
unphysical dipole-like structure.  (The large std in the case of Fourier
components does not mean that for some realizations the Fourier components
would have given a lower rms than the other methods.  Rather, the distribution
was just very skew; a small number of realizations gave a very large rms, but
all 10 realizations gave an rms larger than the average for any of the other
methods.)

The problem with Fourier components is related to small eigenvalues
of matrix $\bA$.
If there exists a map $\ve{m}'$ such that
\beq
   \bP\ve{m}'=\bF\ve{a}'  \label{bad}
\eeq for some $\ve{a}'$, then $\ve{a}'$ is an eigenvector of $\bA$ with zero
eigenvalue. If $\ve{a}$ is a solution of Eq. (\ref{lin3}), then
$\ve{a}+\ve{a}'$ is also a solution. In other words, if we can produce the same
TOD stream both as a combination of the base functions, and by picking it from
a map with our scanning strategy, then it is not possible, without further
information, to tell if this TOD component comes from noise or signal. In
practice zero eigenvalues are unlikely to happen, but already small but
non-vanishing eigenvalues cause similar problems. Eq. (\ref{bad}) then holds
approximatively.

The difficulty of fitting Fourier components is related to the symmetry of the
nominal scanning strategy of {\sc Planck}. Suppose the scanning rings are
regular circles with centers on the ecliptic plane. If we give all cosine and
sine components equal amplitudes $a$ and $b$ such that the resulting function
$a\cos\phi+b\sin\phi$ ($\phi$ runs from $0$ to $2\pi$ around the ring) is at
maximum at the northernmost point and at minimum at the southernmost point (or
vice versa), and coadd a map from this TOD stream, we obtain a meridian
symmetric map for which relation (\ref{bad}) holds. This noise component cannot
be resolved without further information on noise properties. The same problem
also affects fitting 2nd order Legendre polynomials, albeit less severely. We
would expect that other, less symmetric, scanning strategies would not be as
prone to such problems.

We discuss this problem quantitatively in the following.


\begin{table}
\caption[a]{\protect\small Average rms (in $\mu$K) of the residual noise map,
std of rms, and reference rms, for different sets of base functions. The
reference rms is computed from a map from which the reference baseline
functions are removed. The base functions were: Un: uniform baseline, F1: three
Fourier
  modes, L1 (L2): Legendre polynomials up to 1st (2nd) order.
The two first lines ('Un.')
represent the same destriping methods as the first and third lines of Table 1.
}
\begin{center}
\begin{tabular}{llll}
\hline\hline
fit & avg rms$/\mu$K & std of rms/$\mu$K  &  ref. rms/$\mu$K  \\
 \hline
 Un. ($w = 1$)    & 225.162   & $\pm$ 0.072  & 224.117  \\
Un. ($w = 1/n_p$) & 224.444  &  $\pm$ 0.073  & 224.117 \\
F1  & 264.131  &  $\pm$70.460  & 223.621 \\
L1  & 224.463  &  $\pm$ 0.077  & 223.860 \\
L2  & 225.049  &  $\pm$ 0.463  & 223.748 \\
\hline
\end{tabular}
\end{center}
\end{table}

\subsection{Eigenvalue analysis of the destriping problem}

Consider the solution
of Eq. (\ref{lin3}) in light of eigenvalue analysis.

Assume the TOD stream is of the form
\beq
    \ve{y} = \bP\ve{m_0}+\bF\ve{a}_0+\ve{n}
\eeq where $\ve{m}_0$ is the actual map, $\ve{a}_0$ is the ``true'' baseline
vector, and $\ve{n}$ represents the remaining noise component. With these
assumptions, Eq. (\ref{lin3}) becomes
 \beq
    \bF^T\bZ\bF\ve{a} = \bF^T\bZ\bF\ve{a}_0+\bF^T\bZ\ve{n}. \label{lina}
 \eeq

Let $\lambda_i$ and $\ve{u_i}$ be the eigenvalues and corresponding
orthonormal eigenvectors of matrix $\bA=\bF^T\bZ\bF$.
Because $\bA$ is symmetric and non-negative definite,
all eigenvalues are positive or zero,
$\lambda_i\ge0$, and the eigenvectors form a complete orthogonal basis.
The matrix can be presented by its eigenvectors as
\beq
    \bF^T\bZ\bF = \sum_i\lambda_i\ve{u}_i\ve{u}_i^T.
\eeq
We can expand $\ve{a}=\sum_ia_i\ve{u}_i$
and $\ve{a}_0=\sum_ia_i^0\ve{u}_i$
and
\beq
    \bF^T\bZ\ve{n}=\sum_ic_i\ve{u}_i.
\eeq
We substitute these expansions into (\ref{lina}) to find
\beq
    a_i = a_i^0 +c_i/\lambda_i.  \label{acomp}
\eeq
Here it must be understood that this relation holds for components
for which $\lambda_i>0$. For $\lambda_i=0$ one can easily show that $c_i=0$
and $a_i$ remains undefined.

Consider then the statistical properties of coefficients $c_i$.
Since $c_i=\ve{u}_i^T\bF^T\bZ\ve{n}$, we find, assuming
that $\ve{n}$ is white and
$\langle\ve{n}\ve{n}^T\rangle=diag(\sigma^2)$, that
$\langle c_i\rangle=0$ and
\beq
     \langle c_ic_k\rangle
     = \sigma^2\ve{u}_i^T\bF^T\bZ\bF\ve{u}_k
     = \sigma^2\lambda_i\delta_{ik}.
\eeq
Looking back at Eq. (\ref{acomp}) we observe that
\beq
   \langle(a_k-a_k^0)(a_i-a_i^0)\rangle
   = \frac{\sigma^2}{\lambda_i}\delta_{ik}.
\eeq

We see that if one of the eigenvalues is very small, then the inaccuracy in the
corresponding component $a_i$ is very large. Actually, the vanishing
eigenvalues do not pose a problem, since the conjugate gradient algorithm
always sets the corresponding amplitude to zero. Problems are caused by
moderately small, but non-vanishing, eigenvalues. How small a value must be
regarded as zero depends on the floating point accuracy of the computer and on
the convergence criterion one has chosen for the conjugate gradient algorithm.

We have seen that fitting uniform baselines only already gives good
results.There is no advantage in trying to fit additional components which have
a large inaccuracy. Fitting poorly determined components causes more error than
leaving them out entirely. We therefore aim to fit only components that
correspond to a large eigenvalue, and eliminate small eigenvalue components. In
the following we present a practical method to do this. The method presented
does not require full determination of eigenvalues or eigenvectors of matrix
$\bA$, which would be a computationally expensive task.

\subsection{A practical method}
\label{sec:prac}

Consider the following equation:
\beq
    (\bF^T\bZ\bF +\epsilon\bF^T\bF)\ve{a} = \bF^T\bZ\ve{y}  \label{solueps}
\eeq
where $\epsilon$ is a small positive constant.
An eigenvalue analysis similar to that presented above shows that the
solution of Eq. (\ref{solueps}) is related to the solution of (\ref{lin3})
through
\beq
    a_i'=a_i\frac{\lambda_i}{\lambda_i+\epsilon\lambda_{max}},
\eeq
where $\lambda_{max}$ is
the largest possible eigenvalue of $\bA$. With our chosen normalization
$\bF^T\bF=diag(n_b)$ it is equal to the number of samples
on a ring, $\lambda_{max}=n_b$.

The effect of the $\epsilon$ term in Eq. (\ref{solueps}) is to wash out
components with eigenvalues smaller than $\epsilon\lambda_{max}$,
while the large eigenvalue component remains unaffected, as long as
$\epsilon$ is small.
At the limit $\epsilon\rightarrow0$ the solution of Eq. (\ref{solueps})
approaches that of Eq. (\ref{lin3}).
%

We have repeated our computations with this method.
Table 4 presents our results for different values of $\epsilon$. The results
were again obtained using 10 different realizations of the instrument noise,
over which the average and the standard deviation were calculated. We see that,
with values $10^{-6}\le\epsilon\le10^{-3}$, the accuracy of fitting Fourier
components is strongly improved with respect to the $\epsilon=0$ case. Also the
required computation time is reduced.  For 2nd order Legendre polynomials we
also find a clear improvement.

However, the results for multiple base functions are still worse than for
uniform baselines only.

Uniform baselines and 1st order Legendre polynomials, which exhibited no
problems with $\epsilon = 0$, are unaffected with $\epsilon = 10^{-4}$ or less,
but with $\epsilon=10^{-3}$ or larger the accuracy of fitting them begins to
deteriorate.

\begin{figure}[tbh]
\center{\resizebox{\hsize}{!}{\includegraphics{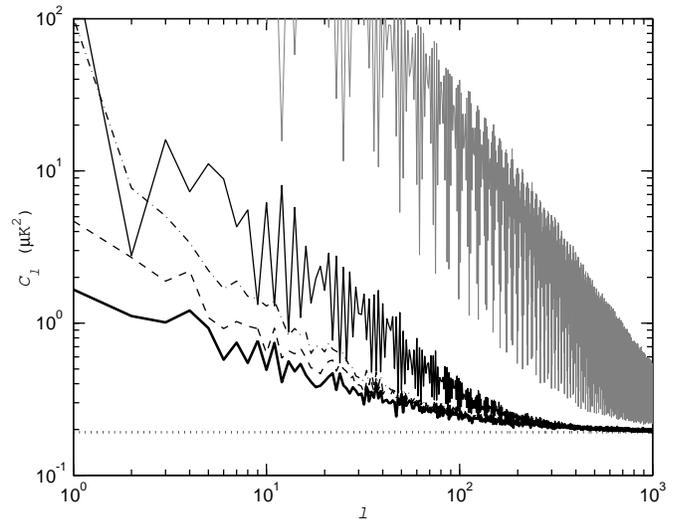}}}
\caption{Same as Fig. \ref{fig:clnd0}, but for the improved method
with $\epsilon=10^{-4}$.
{\it Thick solid} line:
uniform baselines. {\it Thin solid} lower line: three Fourier
components ($q=0,1,2$). {\it Dashed} and {\it dot-dashed} lines:
Legendre polynomials up to 1st and 2nd order.
The corresponding
map rms values are shown in Table 4. }
\label{fig:clndd}
\end{figure}

Fig.~\ref{fig:clndd} shows the residual noise $C_\ell$ spectra for the
$\epsilon = 10^{-4}$ case.

Depending on the number of base functions, the code took 3-5 seconds per
iteration step on one processor of an IBM eServer Cluster 1600 computer. The
total computation time varied between 2 and 30 minutes.


\begin{table}
\caption[a]{\protect\small
Average rms (in $\mu$K) of the residual noise map,
std of rms (middle), and number of iteration steps for
  different sets of base functions and for different values
of $\epsilon$ ($f_k=0.1$ Hz).
The last line gives the reference rms.
Parameter $\epsilon$ is defined in Eq. (\ref{solueps}).
The base functions were: Un: uniform baseline, F1: three Fourier
  modes, L1 (L2): Legendre polynomials up to 1st (2nd) order.
}
\begin{center}
\begin{tabular}{lllll}
\hline\hline
$\epsilon$ & Un.     &  F1     &  L1     &  L2     \\
\hline
 rms \\
\hline
$0$        & 224.444 & 264.131 & 224.463 & 225.049 \\
$10^{-7}$  & 224.444 & 251.648 & 224.463 & 225.047 \\
$10^{-6}$  & 224.444 & 229.675 & 224.463 & 225.025 \\
$10^{-5}$  & 224.444 & 226.034 & 224.463 & 224.866 \\
$10^{-4}$  & 224.444 & 225.640 & 224.463 & 224.563 \\
$10^{-3}$  & 224.463 & 225.438 & 224.502 & 224.678 \\
$10^{-2}$  & 226.192 & 230.141 & 227.691 & 230.396 \\
ref.       & 224.117 & 223.621 & 223.860 & 223.748 \\
\hline
std of rms \\
\hline
$0$        & 0.073 & 70.460 & 0.077 & 0.463 \\
$10^{-7}$  & 0.073 & 48.449 & 0.077 & 0.461 \\
$10^{-6}$  & 0.073 &  7.044 & 0.077 & 0.445 \\
$10^{-5}$  & 0.073 &  0.450 & 0.077 & 0.327 \\
$10^{-4}$  & 0.073 &  0.168 & 0.077 & 0.109 \\
$10^{-3}$  & 0.072 &  0.196 & 0.078 & 0.130 \\
$10^{-2}$  & 0.230 &  0.803 & 0.491 & 1.311 \\
\hline
Iteration steps   \\
\hline
$0$        &  28 & 373 &  39 & 130 \\
$10^{-7}$  &  28 & 369 &  39 & 130 \\
$10^{-6}$  &  28 & 351 &  39 & 130 \\
$10^{-5}$  &  28 & 318 &  39 & 128 \\
$10^{-4}$  &  28 & 166 &  38 & 119 \\
$10^{-3}$  &  27 & 102 &  37 &  95 \\
$10^{-2}$  &  25 &  46 &  31 &  47 \\
\hline
\end{tabular}
\end{center}
\end{table}

\begin{figure}[tbh]
\center{\resizebox{\hsize}{!}{\includegraphics{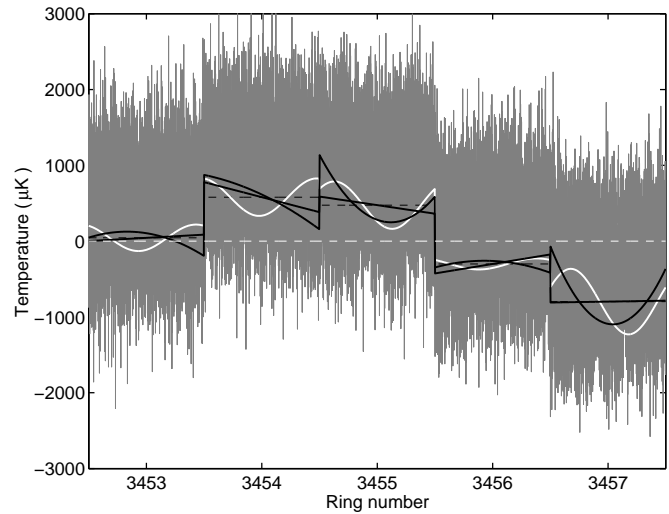}}}
 \caption{Same as Fig. \ref{fig:noise_illustration}, but now with different sets
of base functions.  The amplitudes of the other base functions were much
smaller than the uniform components, so that it would be difficult to see the
small deviation from uniform baselines.  Therefore we have exaggerated the
deviation from the uniform-baseline case by a factor of 5 in this figure. {\it
Dashed} line: uniform baselines. {\it White solid} line: three Fourier
components ($q=0,1,2$). {\it Black solid} lines: Legendre polynomials up to 1st
and 2nd order.
 }
\label{fig:fourier_illustration_bw}
 \end{figure}

To illustrate the use of several base functions we show in
Fig.~\ref{fig:fourier_illustration_bw} the same 5 hour coadded noise TOD as in
Fig.~\ref{fig:noise_illustration}, but now with different sets of base
functions.  Note that the deviations from uniform baselines are exaggerated in
this figure.  The actual amplitudes of the other base functions are much
smaller (by about a factor of 20) than those of the uniform components.

 In order to check how the results depend on the knee frequency, we
repeated our computations with rescaled noise. We took the $1/f$ noise stream,
which was originally generated with $\fk=0.1$ Hz, scaled it by a factor 0.5 or
2, and added white noise with the same variance in all cases. This is
equivalent to changing the knee frequency by a factor of 0.25 or 4. We thus
have results for three knee frequencies: $f_k=0.025$Hz, $\fk=0.1$Hz, and
$\fk=0.4$Hz.

The obtained residual map rms for $f_k=0.4$Hz and $\fk=0.025$Hz are shown in
Tables 5 and 6, for different values of $\epsilon$. The optimal value of
$\epsilon$ seems to depend somewhat on knee frequency, being smaller at higher
knee frequencies.

\begin{figure}[tbh]
\center{\resizebox{\hsize}{!}{\includegraphics{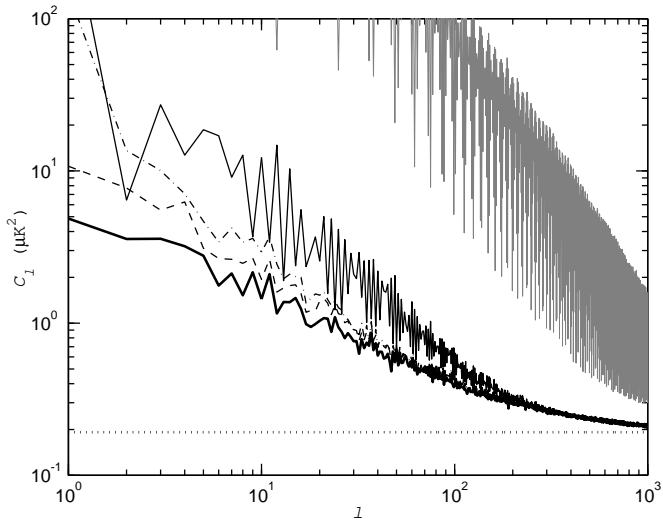}}}
\caption{Same as Fig.~\ref{fig:clndd} but for knee frequency
$\fk=0.4$Hz.} \label{fig:clnhd}
\end{figure}

\begin{figure}[tbh]
\center{\resizebox{\hsize}{!}{\includegraphics{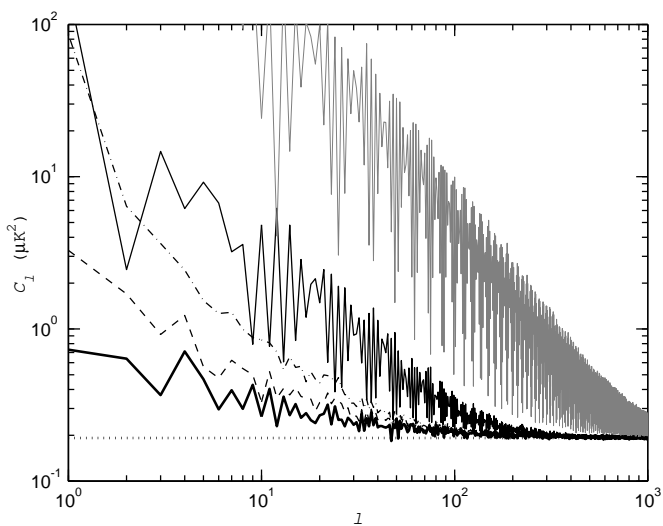}}}
\caption{Same as Fig.~\ref{fig:clndd} but for knee frequency
$\fk=0.025$Hz.} \label{fig:clnld}
\end{figure}

\begin{figure}[tbh]
\center{\resizebox{\hsize}{!}{\includegraphics{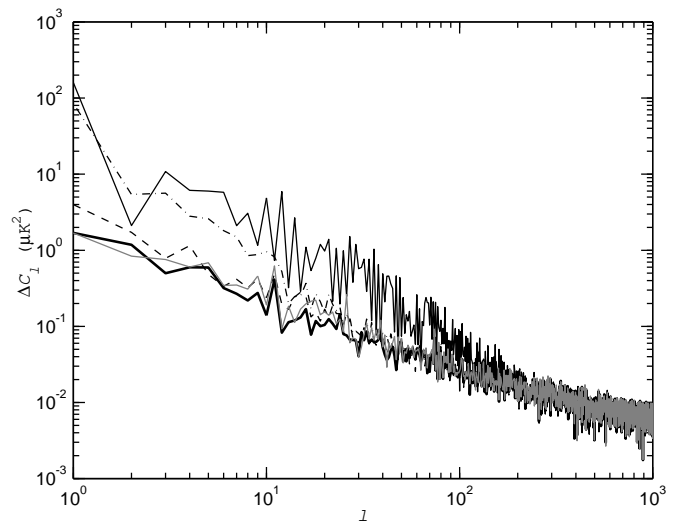}}}
\caption{Standard deviation (std) of the residual noise $C_\ell$ over 10 noise
realizations, for different choices of the base functions ($f_k=0.1$ Hz,
$\epsilon=10^{-4}$). {\it Thick black} line: uniform baselines with $w =
1/n_p$. {\it Grey} line: uniform baselines with $w = 1$. {\it Thin solid} line:
three Fourier components ($q=0,1,2$). {\it Dashed} and {\it dot-dashed} lines:
Legendre polynomials up to 1st and 2nd order. The corresponding map rms and std
of map rms values are shown in Table 4.} \label{fig:clndd_std}
\end{figure}

The $C_\ell$ spectra for $\epsilon=10^{-4}$
for knee frequencies $\fk=0.1\mbox{Hz}$, $\fk=0.4\mbox{Hz}$,
and $\fk=0.025\mbox{Hz}$ are shown in
Figs.~\ref{fig:clndd}, \ref{fig:clnhd}, \ref{fig:clnld},
respectively. The std of the
$C_\ell$ for the $\fk=0.1\mbox{Hz}$ case are shown in Fig.
\ref{fig:clndd_std}.

\begin{figure}[tbh]
\center{\resizebox{\hsize}{!}{\includegraphics{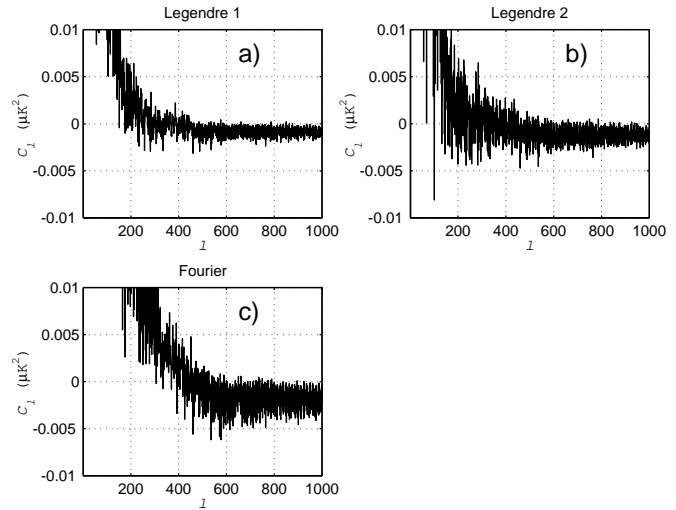}}}
\caption{Differences between $C_\ell$ spectra shown in
Fig.~\ref{fig:clnhd} ($\fk=0.4\mbox{Hz}$). The three panels show
the change in the $C_\ell$ spectrum when fitting Legendre
polynomials up to {\bf a)} 1st or {\bf b)} 2nd order, or {\bf c)}
three Fourier components, instead of uniform baselines only.}
\label{fig:diffhd}
\end{figure}

Looking at the average $C_\ell$ spectra of residual noise
and their std
we see that fitting additional base
functions decreases the accuracy of destriping at low $\ell$.
However the situation for the map rms values in Table 5 seems more
complicated. For $\fk=0.1\mbox{Hz}$ or smaller, fitting additional base
functions does not improve the performance of destriping, but with
$\fk=0.4\mbox{Hz}$ fitting one or two Legendre polynomials besides the
constant baselines decreases the map rms, while Fourier components
still give inferior results. The improvement in the map rms for
$\fk=0.4 \mbox{Hz}$, when fitting Legendre polynomials, comes from the
high multipoles. This can be seen from Fig.~\ref{fig:diffhd},
where we plot the difference between the residual noise $C_\ell$
obtained when fitting Legendre polynomials or Fourier components,
and when fitting uniform baselines only.

We see that increasing the number of base functions improves the high $\ell$
but worsens the low $\ell$ part of the $C_\ell$ spectra. This is true both for
Fourier components and for Legendre polynomials. This trend persists for lower
$\fk$, but the value of $\ell$ above which we get an improvement goes up and
the improvement for those $\ell$ becomes smaller.

Delabrouille (\cite{Delabrouille98}) obtained improved results by fitting
several base functions already with $\fk=0.1\mbox{Hz}$. The difference between
our results and his is probably due to differences in the noise model. While we
assume $P\propto f^{-1}$, as appropriate for LFI radiometers (Seiffert et al.
\cite{Seiffert02}), Delabrouille assumes a noise spectrum of the form $P\propto
f^{-2}$ to account also for possible thermal fluctuations and atmospheric noise
in ground based and balloon borne bolometer experiments. This leads to more
low-frequency noise for a given knee frequency.

The std of the residual noise $C_\ell$ influences the accuracy at
which the $C_\ell$ spectrum of the CMB can be estimated from the
noisy data. From Fig.~\ref{fig:clndd_std} we can see that at low
$\ell$ uniform baselines give the best performance in the sense
that the $C_\ell$ of the residual noise varies the least from one
realization to another. At high $\ell$ there is no clear
difference between the performances of different sets of base
functions.

\begin{table}
\caption[a]{\protect\small Average rms of the residual
noise map (in $\mu$K), for
different sets of base functions and for knee frequency $f_k=0.4$ Hz.
The base functions were: Un: uniform baseline, F1: three Fourier
  modes, L1 (L2): Legendre polynomials up to 1st (2nd) order.
The last line gives the reference rms.
The corresponding
$C_\ell$ spectra for $\epsilon=10^{-4}$ are shown in
Fig.\ref{fig:clnhd}.
}
\begin{center}
\begin{tabular}{lllll}
\hline\hline
$\epsilon$ & Un.     &  F1     &  L1     &  L2     \\
\hline
$0$        & 234.184 & 258.343 & 233.947 & 234.544 \\
$10^{-6}$  & 234.184 & 238.107 & 233.947 & 234.512 \\
$10^{-5}$  & 234.183 & 235.603 & 233.947 & 234.304 \\
$10^{-4}$  & 234.183 & 235.315 & 233.947 & 233.910 \\
$10^{-3}$  & 234.252 & 235.643 & 234.102 & 234.514 \\
ref.       & 233.337 & 231.626 & 232.430 & 232.045 \\
\hline
\end{tabular}
\end{center}
\end{table}


\begin{table}
\caption[a]{\protect\small Same as Table 5, but for
$f_k=0.025$ Hz.
The corresponding
$C_\ell$ spectra for $\epsilon=10^{-4}$ are shown in
Fig.\ref{fig:clnld}.
}
\begin{center}
\begin{tabular}{lllll}
\hline\hline
$\epsilon$ & Un.     &  F1     &  L1     &  L2     \\
\hline
$0$        & 221.943 & 262.149 & 222.030 & 222.597 \\
$10^{-5}$  & 221.943 & 223.588 & 222.030 & 222.433 \\
$10^{-4}$  & 221.943 & 223.173 & 222.030 & 222.163 \\
$10^{-3}$  & 221.948 & 222.825 & 222.038 & 222.149 \\
$10^{-2}$  & 222.385 & 223.676 & 222.835 & 223.565 \\
ref.       & 221.753 & 221.575 & 221.665 & 221.626 \\
\hline
\end{tabular}
\end{center}
\end{table}


\section{Conclusions}
\label{sec:conclu}

We have presented a maximum-likelihood formulation of the destriping approach
to the CMB map-making problem, and a rigorous derivation of the destriping
algorithm, and we have applied it to the case of the {\sc Planck} mission.

We have formulated the method in matrix form, which allows us to
apply the conjugate gradient technique in such a way that we can
handle very large data sets.

We have compared the three different destriping methods, the one
derived here and the other two already presented in the literature,
using simulated {\sc Planck} data (one 100~GHz LFI detector).
The differences between these methods can be expressed in terms of
a weight function, which varies between methods.
This function assigns weights to pixels based on the
number of observations falling on that pixel.

We found that our new method provides some improvement to the method used in
Burigana et al. (\cite{Burigana99}) and Maino et al. (\cite{Maino99,Maino02}).
%
However, our new method was not better than the method given by Delabrouille
(\cite{Delabrouille98}), although he gives only a heuristic justification for
his weight function. The difference between the latter two methods was
insignificantly small, but was systematic. That the maximum-likelihood
derivation did not lead to the optimal method in practice is due to actual
noise properties differing from the idealization used in the derivation. We
recommend using either the weight function derived here ($w=1/n_p$) or the one
given by Delabrouille ($w=1/(n_p-1)$).

We have tested the possibility of improving the accuracy of destriping by
fitting more base functions besides the uniform baseline, but we have found no
systematic improvement in the case of instrumental $1/f$ noise. (Fitting
several base functions may be more beneficial when removing other types of
systematics, i.e. periodic fluctuations induced by thermal instabilities.) The
optimal selection of base functions seems to depend on the actual spectrum of
the 
noise, and on which multipoles one is mainly interested in.
However, the great virtue of the destriping method is its simplicity: it does
not require prior information on the noise spectrum. We lose this advantage if
we incorporate information on the noise spectrum into the method.

\begin{acknowledgements}
This work was supported by the Academy of Finland Antares Space Research
Programme grant no. 51433. TP wishes to thank the V\"{a}is\"{a}l\"{a}
Foundation for financial support. We thank CSC (Finland) and NERSC (U.S.A.) for
computational resources. DM and CB warmly thank all the coauthors of our
``classical destriping code'' and the colleagues that contributed to the
related applications. We acknowledge use of the {\sc cmbfast} code for the
computation of the theoretical CMB angular power spectrum. We gratefully
acknowledge K.~G\'orski and B.~Wandelt for their implementation of the SDE
noise generation method. Some of the results in this paper have been derived
using the HEALPix package (G\'orski et al. \cite{gorski99}).  We warmly thank
the referee for constructive comments on the first version of this paper.

\end{acknowledgements}


\clearpage

\end{document}